\documentclass[english,aps,preprint,showpacs]{revtex4}
\usepackage{mathpazo}
\usepackage[T1]{fontenc}
\usepackage[latin9]{inputenc}
\usepackage{fancyhdr}
\usepackage{graphicx}
\usepackage{esint}

\makeatletter

\providecommand{\tabularnewline}{\\}

\@ifundefined{textcolor}{}
{%
 \definecolor{BLACK}{gray}{0}
 \definecolor{WHITE}{gray}{1}
 \definecolor{RED}{rgb}{1,0,0}
 \definecolor{GREEN}{rgb}{0,1,0}
 \definecolor{BLUE}{rgb}{0,0,1}
 \definecolor{CYAN}{cmyk}{1,0,0,0}
 \definecolor{MAGENTA}{cmyk}{0,1,0,0}
 \definecolor{YELLOW}{cmyk}{0,0,1,0}
 }

\makeatother

\usepackage{babel}

\begin{document}

\title{Information Content of Polarization Measurements}

\author{D G Ireland}

\email{d.ireland@physics.gla.ac.uk}

\affiliation{Department of Physics and Astronomy, University of Glasgow, Glasgow
G12 8QQ, United Kingdom }
\begin{abstract}
Information entropy is applied to the state of knowledge of reaction
amplitudes in pseudoscalar meson photoproduction, and a scheme is
developed that quantifies the information content of a measured set
of polarization observables. It is shown that this definition of information
is a more practical measure of the quality of a set of measured observables
than whether the combination is a mathematically complete set. It
is also shown that when experimental uncertainty is introduced, complete
sets of measurements do not necessarily remove ambiguities, and that
experiments should strive to measure as many observables as practical
in order to extract amplitudes. 
\end{abstract}


\maketitle
\thispagestyle{fancy}

\section{Introduction}

In pseudoscalar meson photoproduction, the reaction is completely
described by four amplitudes that are functions of hadronic mass $W$
and center of mass scattering angle $\theta_{CM}$ (or, equivalently
$s$ and $t$). If one were able to extract these amplitudes (allowing
of course for an overall phase) at $\left\{ W,\cos\theta_{CM} \right\} $ or $\left\{ s,t\right\} $ points,
there is nothing else one could measure that would alter how one could
interpret the physics of the reaction. 

This observation is especially important in the study of the spectrum
of baryon resonances. Despite several decades of investigation, it
is still not clear whether certain states that are predicted by quark
models exist or not. The signatures of any hitherto undiscovered states
must be very subtle, to the extent that they are not readily apparent
from cross section measurements alone. If one could unpick the reaction
amplitudes from suitable observables, that would constitute the most
comprehensive test for models. In the case of establishing $s$-channel resonances, extraction of the four amplitudes may not even be enough. Partial-wave analyses will be required, and these can lead to finite ambiguities that require additional information to resolve. In any event, a potential new physical effect would have
to manifest itself clearly, or be declared unproven.

In order to extract the amplitudes, it is necessary to measure several
polarization observables. In addition to the cross section, there
are three single-spin observables %
\footnote{This departure from the usual conventions is to avoid confusion between
$\Sigma$-particle and $\Sigma$-beam asymmetry, as well as $P$-photon
polarization and $P$-recoil.%
}: $B$ (photon beam asymmetry), $R$ (recoil polarization) and $T$
(target polarization), which can be labelled as ${\cal S}$-type measurements.
There are also four beam-recoil (${\cal BR}$-type), four beam-target
(${\cal BT}$-type) and four recoil-target (${\cal RT}$-type) observables.
All these observables are bilinear combinations of the four reaction
amplitudes, and are not independent. In principle, therefore, it is
not necessary to measure all of them to be able to infer the amplitudes.
As we have now entered an era in which single- and double-polarization
measurements can be made, there exists a real opportunity for progress
in understanding pseudoscalar meson photoproduction reactions, and
for potential discovery of new states.

The problem of finding a minimum set of measurements that allows the
unambiguous extraction of amplitudes was addressed by Barker, Donnachie
and Storrow \cite{Barker1975}). They found that, in addition to the
single polarization set, five more double polarization observables
were needed to remove all ambiguities in the quadrants for each relative
phase angle. More recently, Chiang and Tabakin \cite{Chiang:1996em}
carried out a detailed analysis of the algebra of observables using
Fierz identities, and showed that the selection of just four suitably
chosen double polarization observables was sufficient to remove the
ambiguities. Such sets have been designated as {}``complete'' sets.

The Fierz identity analysis led to a large number of identities among observables. Work by Artru et al.~\cite{Artru:2006xf,Artru20091} extended this by using positivity constraints to derive many \emph{inequalities}. This means that the measurement of a subset of observables places limits on the possible values of the undetermined observables, so the inequalities  provide useful guides to whether the values of experimental data are physical.

Labelling sets of observables as {}``complete'', implies
somehow that one has reached an ultimate state of knowledge. However,
the reality is that all experimental measurements of observables carry
with them a finite uncertainty, so the concept of completeness is
not well defined. One might be tempted to regard this as an experimental
failing, but in practice any experiment has to be performed within
constraints of time and technological feasibility; the experiment
with zero uncertainty can only be accomplished in an infinite time.
The alternative is to embrace experimental uncertainty and include
it in the interpretation of results.

The problem of uncertainty due to noise in communication channels
led Shannon to develop the foundations of information theory \cite{Shannon1948}.
In that seminal work, the concept of entropy was used as a means of
quantifying an amount of information. One can also apply this to measurements.
To introduce the idea with a concrete example, suppose one measured
a quantity $X$ and obtained a measured value $x$ with an uncertainty
$\delta_{x}$. The reporting of this measurement would usually be
in the form $x\pm\delta_{x}$, but this is really shorthand for a
Gaussian probability density function (PDF) $p\left(x\right)$. The
entropy is then\begin{equation}
H=-\int p\left(x\right)\log p\left(x\right)dx,\label{eq:entropy}\end{equation}
which for a Gaussian PDF is $H=\log\sqrt{2\pi e}+\log\delta_{x}$.
If a more accurate measurement were to be made, resulting in a reduced
uncertainty $\delta_{x}^{\prime}$, the gain in information can be
quantified as\[
I=H-H^{\prime}=\log\left(\frac{\delta_{x}}{\delta_{x}^{\prime}}\right).\]

By extending this idea to the uncertainty in the reaction amplitudes,
it is possible to quantify how much information is gained following
the measurement of one or more observables. This article represents
a preliminary study of information entropy as applied to pseudoscalar
meson photoproduction. Section \ref{sec:Measuring-Information} develops
the idea encapsulated by Eq. (\ref{eq:entropy}) for the reaction
amplitudes, and introduces a means of calculating it. In section \ref{sec:Results}
examples of hypothetical measurements are given, which show how the
magnitudes and relative phases of the amplitudes can be determined.
In addition to this, section \ref{sec:Comparison-of-Models} briefly considers
how the information content of measured data can be used as a guide to estimating whether the measurement could in principle reduce uncertainty in derived physical quantities.

\section{\label{sec:Measuring-Information}Measuring Information}

\subsection{Reduced Amplitudes}

A full analysis of reactions will involve measurements over all scattering
angles and cover the mass range of interest. To develop the concept
of information content, however, we restrict ourselves to considering
one region (or {}``bin'') in $\left\{ W,\theta_{CM}\right\} $ space.
The ideas can be straightforwardly extended to include many regions,
since entropies are additive. The issue of whether different experiments
(measuring different observables) have covered the same $\left\{ W,\theta_{CM}\right\} $
space has been avoided.

The choice of basis for amplitudes is arbitrary; information content
is derived from the measured observables, so it cannot depend on the
choice. In this work, the transversity basis is used, where the spin
of the target nucleon and recoiling baryon is projected onto the normal
to the scattering plane, and the linear polarization of the photon
is either normal or parallel to the scattering plane. 

It is assumed that differential cross section measurements have been
performed to a level of accuracy of, say, a few percent, so that further
measurement would be unlikely significantly to improve knowledge of
the amplitudes. The information gain to be studied here is solely
due to an increased accuracy in the knowledge of the polarization
observables. Since all these observables are asymmetries, no generality
is lost if we rescale the amplitudes $b_{i}\rightarrow a_{i}$ such
that\[
a_{i}=\frac{b_{i}}{\sqrt{\left|b_{1}\right|^{2}+\left|b_{2}\right|^{2}+\left|b_{3}\right|^{2}+\left|b_{4}\right|^{2}}},\]
so that the cross section provides an overall scale factor. Applying
this rescaling, we have\begin{equation}
\left|a_{1}\right|^{2}+\left|a_{2}\right|^{2}+\left|a_{3}\right|^{2}+\left|a_{4}\right|^{2}=1.\label{eq:7-sphere}\end{equation}
 Since these reduced amplitudes $a_{i}$ are complex, this represents
the equation of a unit 7-sphere, i.e. the eight numbers that are the
real and imaginary parts are constrained to be on the surface of a
unit hypersphere in 8 dimensions (a unit 8-ball). 

The definitions of the observables in terms of the reduced amplitudes
are given in appendix \ref{sec:Definitions-of-Observables}. One side-effect
of choosing transversity amplitudes is that measurement of the ${\cal S}$-type
observables leads to the extraction of the magnitudes, leaving just
the relative phases to be determined. There is often a tacit assumption
that it is easier to perform single-spin asymmetry measurements. For
that reason many analyses \cite{Barker1975,Chiang:1996em} start from
a point where values of the ${\cal S}$-type observables have been
determined.

\subsection{Entropy}

The entropy associated with the state of knowledge of the amplitudes
is an multidimensional extension of Eq. (\ref{eq:entropy}):\begin{equation}
H=-\int p\left(\left\{ x_{i}\right\} \right)\log p\left(\left\{ x_{i}\right\} \right)d\left\{ x_{i}\right\} ,\label{eq:nd-entropy}\end{equation}
where $\left\{ x_{i}\right\} $ represents the values of the real
and imaginary parts of the amplitudes. Before the measurement of any
polarization observable, there is no knowledge of $\left\{ x_{i}\right\} $,
other than the constraint imposed by Eq. (\ref{eq:7-sphere}). To
encode this as a PDF, we can spread the probability uniformly over
the surface area of the unit 7-sphere to give\[
p\left(\left\{ x_{i}\right\} \right)=\frac{3}{\pi^{4}},\]
which results in a pleasingly simple entropy of \begin{equation}
H_{7-sphere}=-\int\frac{3}{\pi^{4}}\log\left(\frac{3}{\pi^{4}}\right)d\left\{ x_{i}\right\} =4\log\pi-\log3.\label{eq:7-sphere-1}\end{equation}

\begin{figure}
\includegraphics[width=0.8\textwidth]{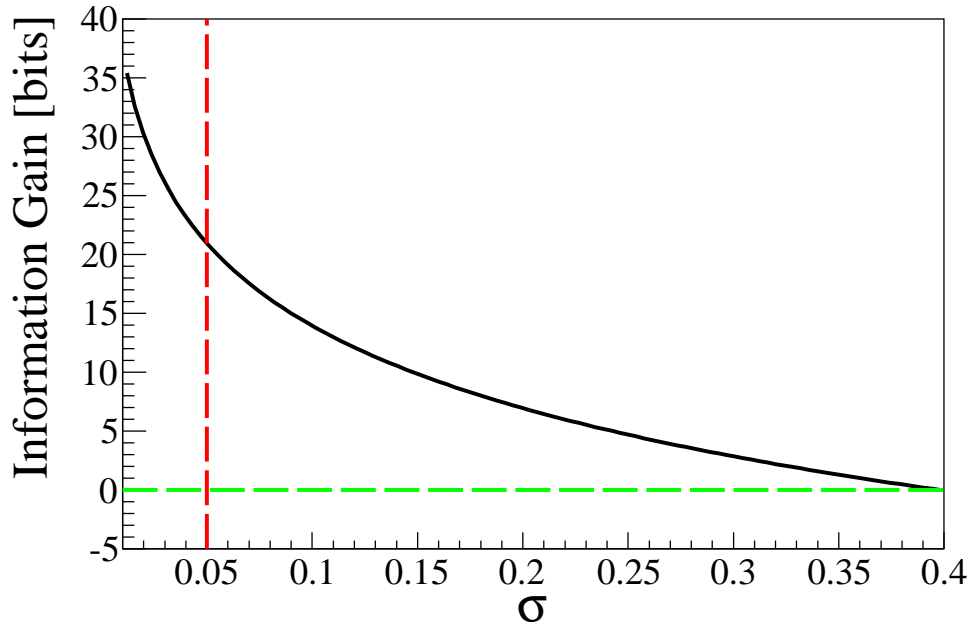}\caption{\label{fig:Rough-guide-to}(Color online) Rough guide to information gain as a function
of the standard deviation $\sigma$ in the real and imaginary parts
of the amplitudes. }

\end{figure}

The act of measurement can be viewed as a compression of this {}``uniform''
PDF into as small a region of $\left\{ x_{i}\right\} $ space as possible.
As a rough example, consider a set of measurements that results in
a multi-dimensional Gaussian PDF in amplitude space. The entropy of
an $n$-dimensional Gaussian is \cite{Shannon1948}\begin{equation}
H_{g}=\frac{n}{2}\log\left(2\pi e\right)+\frac{1}{2}\log\left(\left|c_{ij}\right|\right),\label{eq:entropy-nd-gaussian}\end{equation}
where $\left|c_{ij}\right|$ is the determinant of the covariance
matrix. While the four complex amplitudes have eight numbers in total,
representing real and imaginary parts, all observable quantities are
invariant to the choice on an overall phase angle, so the effective
number of numbers to extract is seven. In this case, a 7-dimensional
Gaussian is used to estimate information gain. The projection of the
Gaussian onto the 7-sphere will induce off-diagonal correlations in
$c_{ij}$, but for simplicity we ignore any correlations and take
the standard deviation in each of the $\left\{ x_{i}\right\} $ to
be the same ($\sigma$, say). The resulting approximate expression
is\begin{equation}
H_{measured}=\frac{7}{2}\log\left(2\pi e\right)+7\log\sigma.\label{eq:entropy-7D-gaussian}\end{equation}
 The gain in information is the difference between this and the initial
uniform PDF over the 7-sphere:\begin{equation}
I=H_{7-sphere}-H_{measured}=4\log\pi-\log3-\frac{7}{2}\log\left(2\pi e\right)-7\log\sigma.\label{eq:info-gain}\end{equation}
A plot of this quantity as function the size of standard deviation
is shown in Fig. (\ref{fig:Rough-guide-to}). The choice of logarithm
base is arbitrary, but for this work we select it to be 2. This means
that the unit of information is the {}``bit'' (i.e. knowing whether
a quantity is 1 or 0). This unit system is convenient for considering
quantities related to polarization; determining whether an asymmetry
is positive or negative is equivalent to a gain of one bit of information,
whereas determining a phase angle quadrant is a gain of two bits.

From Fig. (\ref{fig:Rough-guide-to}) it can be seen that if one wants
to have a measured accuracy of the amplitudes to a value $\sigma=0.05$,
the gain in information is roughly 21 bits (see dashed vertical line
on graph). Attempting to achieve much better accuracy than this from
experiments is not likely to be practical, so we should therefore
regard the 21-bit information gain as a target figure to aim for,
if we want to be able to say that we have extracted amplitudes. Furthermore,
if two models differ by only a few percent in the values of their
amplitudes, it is not reasonable to expect that comparison with data
would ever lead to being able to differentiate them.

\subsection{Numerical calculation of entropies. }

While the calculation sketched out above is a useful rough guide,
when an actual set of observables have been measured, Eq. (\ref{eq:nd-entropy})
will need to be evaluated numerically. The number of dimensions in
this system indicates the use of Monte Carlo techniques, and a simple
implementation of this is as follows.

Sample points are generated randomly in amplitude space with uniform
density on the surface of the unit 7-sphere. The number of points
$N_{0}$ needs to be sufficiently large to minimize Monte Carlo sampling
uncertainty. For each point, the observables are evaluated according
to the algebra of table \ref{tab:Definition-of-observable} in the appendix. The use of random values of amplitudes  was described in \cite{Artru20091} in order to establish,
for combinations of observables, the limits of regions in observable
space that are allowed by postivity constraints, and using this a a guide for deriving inequalities. The present work goes further by not only taking into account these positivity constraints, but also estimating the PDFs of the combinations of observables. One can then simulate
the process of measuring an observable by weighting all the points
by another PDF representing the measured observable. 

In practice, the PDF of an asymmetry is likely to be something like
a beta distribution (or a Gaussian approximation thereof). For illustrative
purposes, however, we can use a simple top-hat function, which for
a single observable is equivalent to reducing the range of values
from $\left[-1,1\right]$ to $\left[r-\delta,r+\delta\right]$, where
$r$ is the measured result with some uncertainty $\pm\delta$. If
the uniform probability density on a multi-dimensional surface $S$
is $p\left(x_{i}\right)d\left\{ x_{i}\right\} =d\left\{ x_{i}\right\} /S$.
The entropy of a uniform distribution in a volume $S$ is then\[
H=-\int\frac{1}{S}\log\left(\frac{1}{S}\right)d\left\{ x_{i}\right\} =\log S,\]
as illustrated by the value for the 7-sphere in Eq. (\ref{eq:7-sphere-1}). 

If the surface is reduced by a cut, say from $S_{0}$ to $S_{1}$,
the probability density will be uniform in $S_{1}$ and zero otherwise,
so the gain in information is simply the log of the ratio of the two
surface areas:\[
I=\log S_{0}-\log S_{1}\]

When cuts representing the measurement of a combination of observables
are imposed, the number of remaining points $N_{1}$ is an estimate
of the remaining volume, so\[
I=\log N_{0}-\log N_{1}.\]
 So in order to gain the 21 bits of information, the surface area
in amplitude space (and hence the number of points) needs to be reduced
by a factor of $2^{21}\approx10^{6}$.

\begin{figure}

\includegraphics[width=0.8\columnwidth]{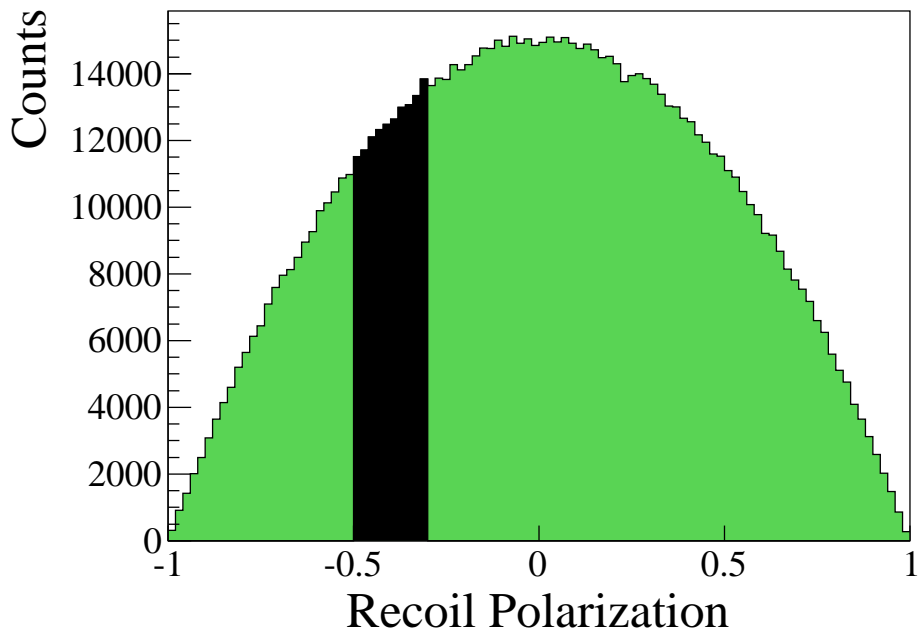}\caption{\label{fig:Distribution-of-values}(Color online) Distribution of values of recoil
polarization from the uniform PDF in amplitude space. Shaded region
represent the possible values remaining after a {}``measurement''. }

\end{figure}

This is best illustrated with a simple example, such as the measurement
of one polarization observable, recoil polarization, say. Figure \ref{fig:Distribution-of-values}
shows in the light shade the distribution of $10^{6}$ points when
sampling is done uniformly in amplitude space. The dark shaded region
shows 126045 points selected when a simulated measurement of $R=-0.4\pm0.1$
is selected. The result is an information gain of $6\log_{2}10-\log_{2}126045=2.988\pm0.003$
bits, where the uncertainty is an estimate of the Monte Carlo error.
So we can expect that a measurement of one polarization observable
to an accuracy of $\pm$10\% will give us about 3 bits of information.

Note that the {}``uncut'' or prior distribution is quadratic in
shape, not only for recoil polarization, but for all observables.
This is a consequence of the observables being bilinear combinations
of the amplitudes.

\section{\label{sec:Results}Simulating Combinations of Measurements}

\subsection{Measuring all ${\cal S}$-type observables}

For the extraction of amplitudes, it is usually assumed that the ${\cal S}$-type
observables ($B$, $R$ and $T$) have to be measured. Let us examine
how much information one gains by making such measurements. 

As shown in \cite{Artru:2006xf,Artru20091}, the constraints among
observables \begin{equation}
\left|T-R\right|\leq1-B;\quad\left|T+R\right|\leq1+B\label{eq:BRT-constraints}\end{equation}
carve out a tetrahedron inside a cube $\left[-1,+1\right]^{3}$ in
$BRT$-space. To approximate a measurement of $B$, $R$ and $T$,
we define a spherical region, of radius $r$, i.e.\[
\left(B-x\right)^{2}+\left(R-y\right)^{2}+\left(T-z\right)^{2}\leq r^{2},\]
where $(x,y,z)$ are the coordinates of the sphere centre. This spherical
cut can be moved to various points within the tetrahedron, and the
effect on the distributions of magnitudes and phases studied. 

\begin{figure}

\includegraphics[width=0.8\textwidth]{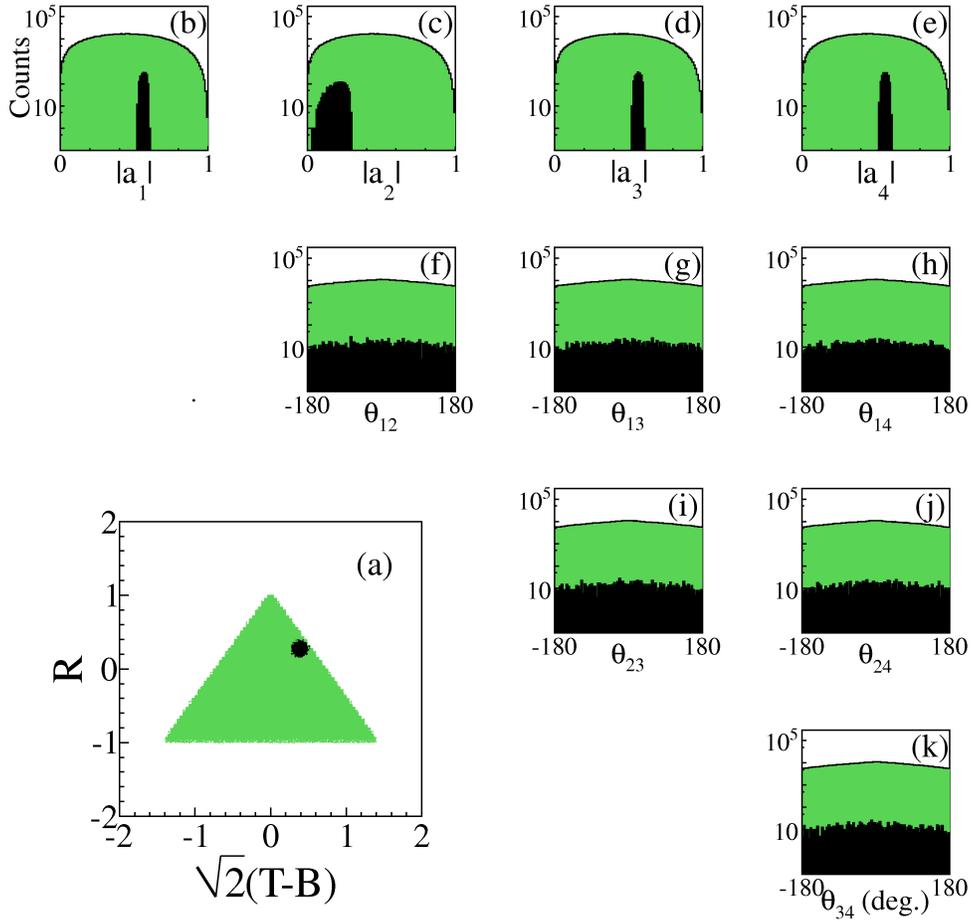}\caption{\label{fig:Bottom-left-panel}(Color online) Light shade - uniform sample of amplitude space; dark shade - region surviving cut. Panel (a) is the projection of BRT tetrahedron, (b)-(e) show the magnitudes of the amplitudes and the other panels are the distributions of relative phase angles (in degrees).}

\end{figure}

A typical example is depicted in Fig. \ref{fig:Bottom-left-panel}.
The bottom left panel shows a projection of the $BRT$ distributions,
which highlights the tetrahedral region. Recall that the points in
the light shaded region have been initially sampled over amplitude
space, so this represents a projection into $BRT$-space, and affirms
the constraints defined by Eq. (\ref{eq:BRT-constraints}). The points
in the dark sphere are those selected by the choice of cut region.
The radius of the spherical cut is 0.1, which is equivalent in information
gain to a measured accuracy in each observable of better that $\pm0.05$
(see later). It is unlikely, when statistical and systematic uncertainties
are taken into account, that experiments will be able to determine
observables to much greater accuracy than this.

In the example of Fig. \ref{fig:Bottom-left-panel}, the spherical
cut is just touching the midpoint of one of the tetrahedron faces.
The top row shows the magnitudes of the amplitudes, and it is clear
that values for each one can now be estimated. Note, however, that
there is much greater uncertainty in $\left|a_{2}\right|$ than in
the other ones. 

The relative phase angles are displayed in the remaining panels. While
only three relative angles are independent, all six possibilities
are shown. This is because, for situations in which the magnitudes
of two amplitudes $a_{i}$ and $a_{j}$ are almost equal (as in this
case), very small uncertainties in the relative phase of the two amplitudes
with respect to a third ($\theta_{ik}$ and $\theta_{jk}$) could
lead to very large uncertainties in their relative phase $\theta_{ij}$.
It is to be expected that there should be no relative phase information
for transversity amplitudes if only ${\cal S}$-type measurements
are made, and this is apparent from the distributions in Fig. \ref{fig:Bottom-left-panel}.
The observed increase towards $\theta_{ij}=0^{\circ}$ is due to the
fact that the relative angles are formed from the difference of two
uniform random variables.

By examining the variations in the distributions of magnitudes and
phases for different positions in the $BRT$ tetrahedron, one can
deduce some general heuristics governing the relation between what
we shall call a $BRT$ measurement and the magnitudes $\left|a_{i}\right|$.
These are listed in table \ref{tab:Guide-to-relative}.

\begin{table}
\begin{tabular}{l|l}
Position in $BRT$ tetrahedron & Magnitude information\tabularnewline
\hline
Center & All magnitudes equal\tabularnewline
Mid-point of face & One magnitude small, others large and equal\tabularnewline
Mid-point of edge & Two magnitudes small and equal, other two large and equal\tabularnewline
Vertex & One magnitude large, others small and equal\tabularnewline
\end{tabular}

\caption{\label{tab:Guide-to-relative}Guide to relative size of magnitudes
for various positions within the $BRT$ tetrahedron}

\end{table}

Returning to the information gain from a $BRT$ measurement, if we
assume that the sampled points in amplitude space project into a uniformly
dense $BRT$ tetrahedron, the entropy before a measurement is\[
H_{tetra}=\log8-\log3,\]
i.e. the volume is a third of the cube $\left[-1,+1\right]^{3}$.
A 3D gaussian, with symmetric widths $\sigma$ has entropy\begin{equation}
H_{3DGaussian}=\frac{3}{2}\log\left(2\pi e\right)+3\log\sigma,\label{eq:3D-Gaussian}\end{equation}
from Eq. (\ref{eq:entropy-nd-gaussian}). To establish an equivalent
spherical cut, one can use the entropy of a sphere of radius $r$
(inside tetrahedron),\begin{equation}
H_{sphere}=\log\left(\frac{4}{3}\pi r^{3}\right),\label{eq:sphere}\end{equation}
and equate Eq. (\ref{eq:3D-Gaussian}) and Eq. (\ref{eq:sphere})
to establish a relationship between $r$ and $\sigma$:\[
\log r-\log\sigma=\frac{1}{2}\log\left(2\pi e\right)-\frac{1}{3}\log\left(\frac{4}{3}\pi\right),\]
from which we have $r\approx2.564\sigma$, so a spherical cut of radius
0.1 is equivalent to Gaussian errors on $B$, $R$ and $T$ with $\sigma=0.039$.
Using these figures, the predicted information gain is 9.31 bits for
any position of the spherical cut within the tetrahedron. For the
case depicted in Fig. \ref{fig:Bottom-left-panel}, the estimated
information gain is 9.28 $\pm$ 0.02. 

It can be readily demonstrated that the estimates of information gain
are equal to the predicted value of 9.31 (to within sampling errors)
for all the classes of position listed in table \ref{tab:Guide-to-relative},
hence verifying the assumption that the $BRT$ tetrahedron is uniformly
sampled. So for real experiments, knowing the \emph{uncertainties}
of the measured values of $B$, $R$ and $T$ allows a calculation
of information gain, irrespective of the \emph{size} of the measured
values.

\subsection{Towards Extraction of Amplitudes}

Compared to the original guideline of 21 bits, we can see that the
measurement of just ${\cal S}$-type observables leaves a lot of information
to be gained. With just under 10 bits, the magnitudes can be determined
to roughly 10\% accuracy, but to determine relative phases to better
than, say, a 16th ($=2^{-4}$) of the full angular range will require
an additional 4 bits for each one. Adding this information together
brings us to 22 bits, one greater than the original estimate. One
might imagine that measurement of an additional four double polarization
observables would now be sufficient, given that individual measurements
can gain about 3 bits (see section \ref{sec:Measuring-Information}).
However, the complicated relations among observables now conspire
against this. 

Chiang and Tabakin \cite{Chiang:1996em} systematically listed the
possible combinations of observables that would lead to a {}``complete''
set; there are a large number of them. They took one example set which
showed a counter-example to the claim in \cite{Barker1975} that completeness
could only be attained if five observables are measured, of which
four should not be from the same ${\cal BR}$-, ${\cal BT}$- or ${\cal RT}$-set.
In that example, $F$, $G$ and $L_{x}$ were taken to be measured,
and whereas Ref. \cite{Barker1975} claimed that $E$ and $L_{z}$
were needed, Ref. \cite{Chiang:1996em} asserted that only $T_{x}$
was necessary.

Using the scheme already outlined, we may examine what happens when
simulated measurements are made of the same sets of observables. The
$BRT$ measurements are all assumed to have been made, but to study
whether the results for information gain depend on the measure values,
four possible cases of position in the measured $BRT$ tetrahedron
have been used: center, mid-face, mid-edge and vertex. They give a
representative sample of all possible cases, and due to the tetrahedral
symmetry only one mid-face, mid-edge and vertex needs to be considered.
For each of the four cases, $10^{5}$ events were generated within
the defined spherical sub-region of the $BRT$ tetrahedron. These
were selected by rejection from an intial uniform sample over amplitude
space (the 7-sphere). In order to simulate possible measurements of
$F$, $G$ and $L_{x}$, a $\pm$0.1 cut on the generated points around
a central value of each observable was imposed. The central values
are shown for each case in table \ref{tab:Results-of-simulated}.
Relatively large values were chosen for clarity of illustration, and
note that the same values of $F$, $G$ and $L_{x}$ could not be
used for each case because of the interdependency of these observables
with the chosen $BRT$ values.

For each set of $BRT$ values, three cases were studied for combinations
of further measurements: $T_{x}$ only (choice of Ref. \cite{Chiang:1996em}),
$E$ and $L_{z}$ (choice of Ref. \cite{Barker1975}) and $T_{x}$,
$E$ and $L_{z}$. Again, a $\pm$0.1 cut on the generated points
around a central value of each observable is applied. The results
for information gain are shown in the penultimate column of table
\ref{tab:Results-of-simulated}.

\begin{table}

\begin{tabular}{|c|c|c|c|c|c|c|c|c|}
\hline 
$BRT$ position & $F$ & $G$ & $L_{x}$ & $E$ & $L_{z}$ & $T_{x}$ & Information (bits) & Ambiguity?\tabularnewline
\hline
 & 0.4 & 0.4 & 0.3 & - & - & 0.7 & 11.4 $\pm$ 0.2 & Y\tabularnewline
Center & 0.4 & 0.4 & 0.3 & 0.3 & 0.3 & - & 12.7 $\pm$ 0.3 & N\tabularnewline
 & 0.4 & 0.4 & 0.3 & 0.3 & 0.3 & 0.7 & 13.2 $\pm$ 0.3 & N\tabularnewline
\hline 
 & 0.4 & -0.4 & 0.4 & - & - & 0.4 & 11.1 $\pm$ 0.1 & Y\tabularnewline
Mid-Face & 0.4 & -0.4 & 0.4 & 0.7 & -0.7 & - & 12.0 $\pm$ 0.2 & N\tabularnewline
 & 0.4 & -0.4 & 0.4 & 0.7 & -0.7 & 0.4 & 12.7 $\pm$ 0.3 & N\tabularnewline
\hline 
 & 0.4 & 0.4 & 0.4 & - & - & -0.7 & 12.4 $\pm$ 0.2 & N\tabularnewline
Mid-Edge & 0.4 & 0.4 & 0.4 & 0.2 & -0.7 & - & 13.6 $\pm$ 0.4 & N\tabularnewline
 & 0.4 & 0.4 & 0.4 & 0.2 & -0.7 & -0.7 & 13.6 $\pm$ 0.4 & N\tabularnewline
\hline 
 & 0.4 & 0.4 & 0.4 & - & - & 0.3 & 8.8 $\pm$ 0.1 & Y\tabularnewline
Vertex & 0.4 & 0.4 & 0.4 & 0.3 & 0.2 & - & 11.1 $\pm$ 0.1 & N{*}\tabularnewline
 & 0.4 & 0.4 & 0.4 & 0.3 & 0.2 & 0.3 & 11.5 $\pm$ 0.2 & N{*}\tabularnewline
\hline
\end{tabular}\caption{\label{tab:Results-of-simulated}Results of simulated measurements
for different combinations of observables. The values of each observable
are all defined with a $\pm$0.1 cut.}

\end{table}

Several points are apparent from the results displayed. It is clear
that the more measurements that are made, the more information that
is gained. It is also clear that the information gain is dependent
on the assumed measured $BRT$ values. Recall that the information
gain obtained when measuring \emph{only} $BRT$ values was independent
of position in the $BRT$ tetrahedron; the difference is again due
to the interdependency among observables. When the information gain
is greater that 13, the number of points surviving the cuts is 10
or less, so the estimates are of limited accuracy. 

\begin{figure}

\includegraphics[width=0.8\textwidth]{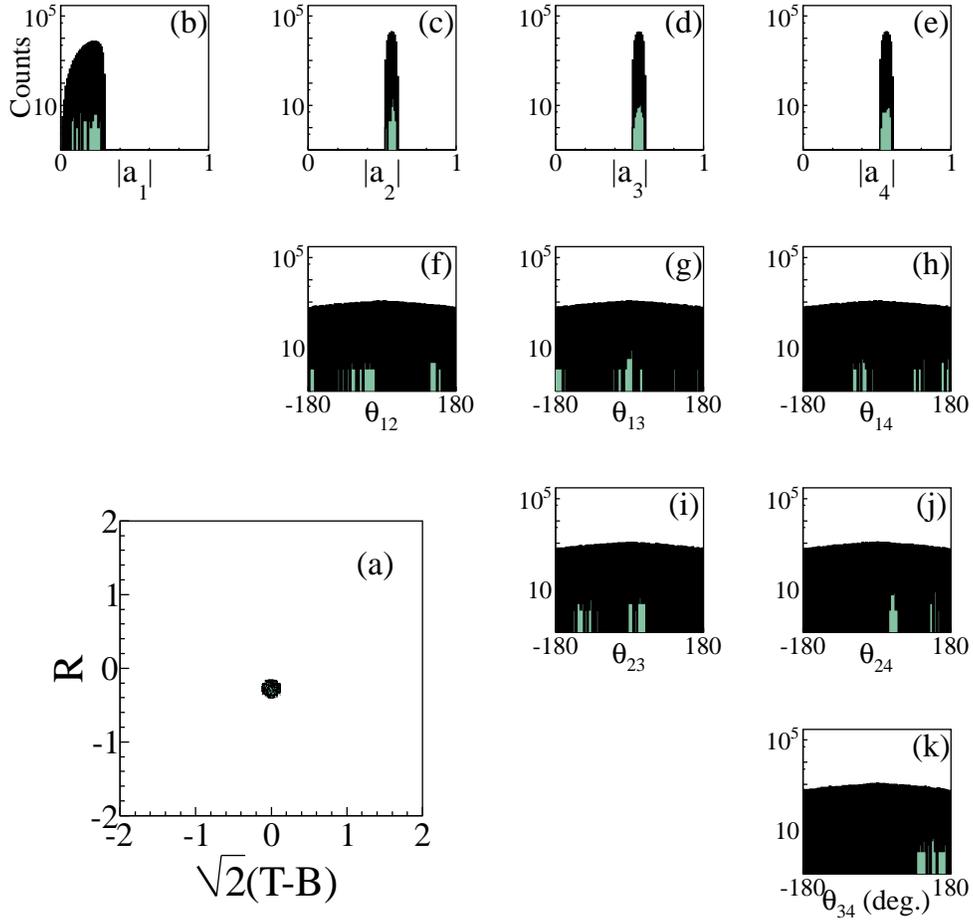}\caption{\label{fig:Example-of-residual}
(Color online) Example of residual ambiguity in relative phases after measurement of set $F$, $G$, $L_{x}$ and $T_{x}$. Light shade - uniform sample of amplitude space; Dark shade - region within BRT tetrahedron; Light shade - points surviving cuts in $F$, $G$, $L_{x}$ and $T_{x}$. Labels (a)-(k) are as in Fig. \ref{fig:Bottom-left-panel}.}

\end{figure}

All the cases of combinations of observables that are listed in table
\ref{tab:Results-of-simulated} have previously been proved to result
in mathematically complete sets. With the introduction of simulated
experimental uncertainty, however, this can no longer be taken to
be adequate. The last column of the table (headed {}``Ambiguity?'')
indicates whether there are identifiable, unambiguous values of both
magnitudes and relative phases of the amplitudes. The mid-face case,
where $T_{x}$ only has been measured in addition to the common set
of observables, is illustrated in Fig. \ref{fig:Example-of-residual}.
Despite the few surviving points, it is fairly clear that there are
no three relative phase angles that have a single cluster of points,
and so an unambiguous extraction of amplitudes would not be possible.

For the cases listed in table \ref{tab:Results-of-simulated} with
N{*} for ambiguity, this indicates that while there is just one identifiable
cluster of points in the distributions of relative phases, the spread
in possible points is greater than 10\% of the full angular range;
i.e. there may be no quadrant ambiguity, but there remains a considerable
uncertainty. 

It appears, from this very small sample of possible outcomes, that
for measurement of double polarization observables an information
gain of about 12 bits is required. Combining this number with that
from the measurement of $BRT$ ($\sim$10 bits), this leads us to
a crude, but very helpful conclusion: only when the total information
gain from polarization observables is greater than about 21 bits should
it be possible to extract amplitudes from experimental measurements.
This condition will apply irrespective of which particular set of
observables have been measured, since information gain is simply a
measure how of much one has compressed the original uniform PDF in
amplitude space. This number is also in line with the crude calculation
given in Eq. (\ref{eq:info-gain}), where the real and imaginary parts
of the amplitudes were assumed to be extracted to an accuracy of 0.05.

The scheme outlined above uses {}``cuts'' in the space of possible
observables to simulate the act of measurement, and the reduced observable
space is projected back into amplitude space to calculate the associated
entropy. This is a crude, but effective, means of relating the observable-space
PDF to the amplitude-space PDF. To apply the idea of information gain
to the results of actual experiments, this scheme will have to be
modified. When the measurement of a set of observables is made, the
result will be an approximately multi-dimensional Gaussian PDF over
the range of those observables. A PDF in amplitude space can be constructed
by sampling uniformly over all amplitude space, calculating the value
of the observables for each sample point then weighting them with
the values of the experimentally determined observable PDF. The resulting
amplitude PDF can be made arbitrarily accurate, depending on the number
of sample points, but for calculating information gains of 21 bits,
${\cal O}\left(10^{7}\right)$ points may be needed. 

One final comment related to practical experiments is in order. It
is clear that for extraction of amplitudes, it is essential to be
able to polarize photon beams and targets, and to detect recoil polarization.
Given that all three components of the reaction require this technological
effort, the most obvious strategy is to worry less about which combination
of observables to measure, and more about trying to measure as many
as possible, with as great an accuracy as possible. The theoretical
work in Refs. \cite{Barker1975,Chiang:1996em} is, however, still
a useful guide to selecting the combinations of observables that will
most efficiently lead to an information gain of 21 bits. The information
measure (\ref{eq:nd-entropy}) can be used in the design of experiments
to provide an estimate of the degree of accuracy (and hence the integrated
luminosity) required for amplitude extraction.

\section{\label{sec:Comparison-of-Models}Comparison of Models with Data}

Having established how to estimate the quantity of information contained
in measured data, can the measured data be used to extract information
about the parameters of an individual model?

An individual model will depend on some input parameters, $\xi$,
say (e.g. coupling constants). Quite often, the comparison of model
calculations to data is used to extract {}``best fit'' values for
the input parameters, $\xi^{\star}$. We can use the information gain
for measured data to tell whether a fit to the new data will yield
an improved knowledge of input parameters, compared to prior information.
Prior to a fit procedure, knowledge about the possible values $\xi$
will be encoded in a PDF $p\left(\xi,M\right)$, where $M$ is there
as a reminder that this quantity depends on a model. The amplitude
PDF of the model, given a specified set of input parameters is $p\left(x_{i}\mid\xi,M\right)$,
where as before $x_{i}$ represents the real and imaginary parts of
the amplitudes. The total prior PDF of the model is an integral over
the range of input parameters\[
p\left(x_{i}\mid M\right)=\int p\left(x_{i}\mid\xi,M\right)p\left(\xi,M\right)d\xi,\]
so a model entropy $H_{model}$ can be evaluated by plugging the model
prior PDF into Eq. (\ref{eq:nd-entropy}). Then only if $H_{measured}\ll H_{model}$
is there likely to be a significant improvement in the knowledge of
the input parameters when the model is fitted to the data.

\section{Conclusions}

In this article a measure of information content, based on Shannon
entropy, was introduced and has been applied to measurement of polarization
observables in pseudo-scalar meson photoproduction. Using the uncertainties
in the measurements, the information entropy of the four amplitudes
can be calculated. It is assumed that a suitably accurate determination
of the cross-section has been made, which gives an overall scale factor
to the amplitudes. 

An important finding is that, when allowing for a realistic but small
measurement uncertainty, measuring only a mathematically complete
set of observables is not enough to guarantee the extraction of amplitudes.
Instead, a rule of thumb, based on quantifying information gain of
about 21 bits for each point in $\left\{ W,\theta_{CM}\right\} $,
is likely to be a more robust guide.

An extension of the work presented here is likely to be applicable
to other reactions in which information content could determine whether
measurements will be adequate to extract physically meaningful results.
Examples such as the extraction of generalized parton distributions
from DVCS-like asymmetries, or inferring the details of the nucleon-nucleon
interaction from the database of scattering observables may be fruitful
areas of investigation.
\begin{acknowledgments}
This work was supported by the United Kingdom's Science and Technology
Facilities Council.
\end{acknowledgments}

%\bibliographystyle{h-physrev}
%\bibliography{/home/dave/Dropbox/Articles/database}

\appendix

\section{\label{sec:Definitions-of-Observables}Definitions of Observables}

In this work, transversity amplitudes are used, although none of the
results depend on using this basis. The \emph{transversity basis}
refers to the projection of the recoil or target spin along the normal
to the scattering plane. The four amplitudes for pseudo-scalar meson
photoproduction are\[
\begin{array}{ll}
b_{1}=\left\langle +\mid{\cal M}\mid+\perp\right\rangle ;\quad & b_{2}=\left\langle -\mid{\cal M}\mid-\perp\right\rangle ;\\
b_{3}=\left\langle +\mid{\cal M}\mid-\parallel\right\rangle ;\quad & b_{4}=\left\langle -\mid{\cal M}\mid+\parallel\right\rangle .\end{array}\]
Here, The labels $\parallel$ and $\perp$ stand for the photon $E$-vector
(linear polarisation) parallel to and normal to the scattering plane
respectively. In order to deal only with asymmetries these amplitudes
are divided by the cross-section to give\[
a_{i}=\frac{b_{i}}{\sqrt{\left|b_{1}\right|^{2}+\left|b_{2}\right|^{2}+\left|b_{3}\right|^{2}+\left|b_{4}\right|^{2}}}\]
The experimentally measured observables are then related to the $a_{i}$
as shown in table \ref{tab:Definition-of-observable}. Whereas the
double poalrization observables are denoted in a conventional way,
$R$, $B$ and $T$ have been chosen to represent Recoil, Beam and
Target asymmetries. 

\begin{table}

\begin{tabular}{ccc}
Observable & Type & Amplitude Combination\tabularnewline
\hline
\hline 
$R$ & Single & $\left|a_{1}\right|^{2}-\left|a_{2}\right|^{2}+\left|a_{3}\right|^{2}-\left|a_{4}\right|^{2}$\tabularnewline
$B$ &  & $\left|a_{1}\right|^{2}+\left|a_{2}\right|^{2}-\left|a_{3}\right|^{2}-\left|a_{4}\right|^{2}$\tabularnewline
$T$ &  & $\left|a_{1}\right|^{2}-\left|a_{2}\right|^{2}-\left|a_{3}\right|^{2}+\left|a_{4}\right|^{2}$\tabularnewline
\hline 
$E$ & Beam-Target & $2\Re\left(a_{1}a_{3}^{\star}+a_{2}a_{4}^{\star}\right)$\tabularnewline
$F$ &  & $2\Im\left(a_{1}a_{3}^{\star}-a_{2}a_{4}^{\star}\right)$\tabularnewline
$G$ &  & $2\Im\left(a_{1}a_{3}^{\star}+a_{2}a_{4}^{\star}\right)$\tabularnewline
$H$ &  & $-2\Re\left(a_{1}a_{3}^{\star}-a_{2}a_{4}^{\star}\right)$\tabularnewline
\hline 
$C_{x}$ & Beam-Recoil & $-2\Im\left(a_{1}a_{4}^{\star}-a_{2}a_{3}^{\star}\right)$\tabularnewline
$C_{z}$ &  & $2\Re\left(a_{1}a_{4}^{\star}+a_{2}a_{3}^{\star}\right)$\tabularnewline
$O_{x}$ &  & $2\Re\left(a_{1}a_{4}^{\star}-a_{2}a_{3}^{\star}\right)$\tabularnewline
$O_{z}$ &  & $2\Im\left(a_{1}a_{4}^{\star}+a_{2}a_{3}^{\star}\right)$\tabularnewline
\hline 
$T_{x}$ & Target-Recoil & $2\Re\left(a_{1}a_{2}^{\star}-a_{3}a_{4}^{\star}\right)$\tabularnewline
$T_{z}$ &  & $2\Im\left(a_{1}a_{2}^{\star}-a_{3}a_{4}^{\star}\right)$\tabularnewline
$L_{x}$ &  & $-2\Im\left(a_{1}a_{2}^{\star}+a_{3}a_{4}^{\star}\right)$\tabularnewline
$L_{z}$ &  & $2\Re\left(a_{1}a_{2}^{\star}+a_{3}a_{4}^{\star}\right)$\tabularnewline
\hline
\end{tabular}\caption{\label{tab:Definition-of-observable}Definition of observable quantities
in terms of the (scaled) amplitudes.}

\end{table}

\end{document}